\documentclass[twocolumn,amssymb, nobibnotes, aps, pre]{revtex4-1}

\usepackage{graphicx}

\begin{document}

\title{A time-periodic mechanical analog of the quantum harmonic oscillator}

\author{Arnaud Lazarus}%
\email[Corresponding author: ]{arnaud.lazarus@upmc.fr}
\affiliation{Sorbonne Universit\'es, UPMC Univ Paris 06, CNRS, UMR 7190, Institut Jean Le Rond d'Alembert, F-75005, Paris, France}
\date{May 30, 2017}%

\begin{abstract}   
We theoretically investigate the stability and linear oscillatory behavior of a naturally unstable particle whose potential energy is harmonically modulated. We find this fundamental dynamical system is analogous in time to a quantum harmonic oscillator. In a certain modulation limit, a.k.a. the Kapitza regime, the modulated oscillator can behave like an effective classic harmonic oscillator. But in the overlooked opposite limit, the stable modes of vibrations are quantized in the modulation parameter space. By analogy with the statistical interpretation of quantum physics, those modes can be characterized by the time-energy uncertainty relation of a quantum harmonic oscillator. Reducing the almost-periodic vibrational modes of the particle to their periodic eigenfunctions, one can transform the original equation of motion to a dimensionless Schr\"odinger stationary wave equation with a harmonic potential. This reduction process introduces two features reminiscent of the quantum realm: a wave-particle duality and a loss of causality that could legitimate a statistical interpretation of the computed eigenfunctions. These results shed new light on periodically time-varying linear dynamical systems and open an original path in the recently revived field of quantum mechanical analogs.
%
\end{abstract}

\maketitle

\section{Introduction}
Modal analysis is a linear perturbation method that allows to characterize the local oscillatory and stability behavior of stationary states of dynamical systems \cite{Guckenheimer1983,Strogatz2001}. It is used in various area of physics, from molecular vibrational frequencies \cite{Nakamoto1986} to the stability of engineered structures \cite{Nayfeh2008}. Reduced to a single dimension in space, this concept is represented by the archetypal example of a mass moving in a local quadratic potential energy whose governing equation is a classic linear homogenous Ordinary Differential Equation (ODE) with initial conditions. In the case of a perturbed equilibrium, i.e. for a constant potential in time, two qualitative behaviors exist: the mass is either stable, harmonically oscillating in a potential well (this case is the classic harmonic oscillator) or unstable, exponentially diverging on a potential hill.

A less constrained situation eventually occurs when the potential energy of the perturbed stationary state is free to periodically vary with time \cite{berge1984,bolotin1964}, i.e. when the motion of the mass is mathematically governed by a linear homogenous ODE with periodically time-varying coefficients. This generalized framework explains a broader class of physical problems from parametric oscillators \cite{Turner1998,Amin2012} to the emergence of Faraday waves \cite{Kumar1994,Melo1994,Engels2007} or the motion of the lunar perigee \cite{Hill1886,Poincare1886}. Again, two different scenarios should be considered whether the system is fundamentally stable or not. In a periodically time-varying potential well, the quasi-periodically oscillating particle eventually destabilizes for certain regions in the modulation parameters space, a.k.a. Mathieu's tongues \cite{Lazarus2010,Bastidas2012}. On a modulated potential hill, there is an asymptotic limit for which the modulation parameters allow to stabilize the fundamentally unstable mass \cite{Kapitsa1951,Grescho1970}. Kapitza's pendulum, the inverted pendulum in which the pivot point vibrates in the vertical direction, is such a system \cite{Stephenson1908,Acheson1993}.  

Although simplistic, this oscillatory vision allows for the description of an outstanding number of physical systems, with the notorious exception of quantum phenomena whose theoretical framework has departed from classical physics in the early $20^{th}$ century \cite{Messiah1961}. In quantum mechanics, a particle modeled by a harmonic oscillator is no more governed by a deterministic linear ODE with any possible initial conditions or mechanical energy, but by Schr\"odinger's equation, a space-varying linear ODE with boundary conditions whose discrete eigensolutions or wavefunctions, associated with a particular energy, represent the probability to observe the particle at a given location. Schr\"odinger's equation undeniably models light-matter interactions at quantum scales but its mathematics sheds no light on the underlying physics which therefore often appears enigmatic \cite{Merali2015}. One hope to pave the way to a realistic interpretation and understanding of quantum physics lies in the use of mechanical analogs. For example, the so called pilot-wave interpretation, first introduced by De Broglie and Bohm \cite{DeBroglie1924,Bohm1952}, have regained considerable attention due to the discovery of a hydraulic quantum analog that consists in a bouncing droplet "walking" on a vertically vibrated bath \cite{Protiere2006,Couder2006,Bush2015,Bush2015b}. By varying the experimental setup, this new framework has already allowed to describe quantum tunneling \cite{Eddi2009}, quantization of classical orbits \cite{Fort2010} and the quantum harmonic oscillator \cite{Perrard2014} in an analogous fashion. Although promising, this interpretation has the disadvantage that the quantum analogy is achieved thanks to highly nonlinear interactions between a wave and a particle, a rather complex model when Schr\"odinger's equation reduces to a $1D$ linear ODE in the case of the quantum harmonic oscillator \cite{Messiah1961}. Therefore, the quest for new and somehow simpler quantum mechanical analogs is still open.

Here, we present a numerical and theoretical study of an overlooked $1D$ time-varying linear oscillator that, in many ways, is analogous to the quantum harmonic oscillator. The described dynamical system is the aforementioned naturally collapsing mass that can be stabilized by the harmonic modulation of its quadratic potential energy, a.k.a. the linearized Kapitza pendulum \cite{Stephenson1908,Kapitsa1951,Acheson1993}.
In the next section, we present the system under study and the numerical tools we use to perform its modal and stability analysis. We then recall the Kapitza asymptotic regime where the modulation periodicity is much smaller than the collapsing time and where the oscillating mass can be modeled as an effective classic harmonic oscillator. In Section 3, we show that in the overlooked opposite asymptotic limit, the stability of the mass gets quantized in various vibrational modes in the modulation parameter space. For those modes, we find that the position and velocity of the mass verify some relations that are similar to the uncertainty principle of the quantum harmonic oscillator \cite{Messiah1961}. In Section 4, we demonstrate that the time-varying ODE governing the stable vibrational modes of the particle can be reduced into the dimensionless Schr\"odinger stationary wave equation of a quantum harmonic oscillator. Finally in Section 5, we discuss various questions that arise from this new fundamental mechanism from the eventualities of experimental realizations to the limitations of the quantum analogy.
Besides its contribution to the field of mechanical quantum analogs, this work would improve our understanding of periodically time varying systems (governed by Floquet theory) and the duality between initial value problem in time and boundary problem in space, that is at the heart of the recently unveiled discrete time crystals \cite{Wilczek2012,Zhang2017,Choi2017,Sacha2017}.

\begin{figure}[!h]
\begin{center}
\includegraphics[width=1\columnwidth]{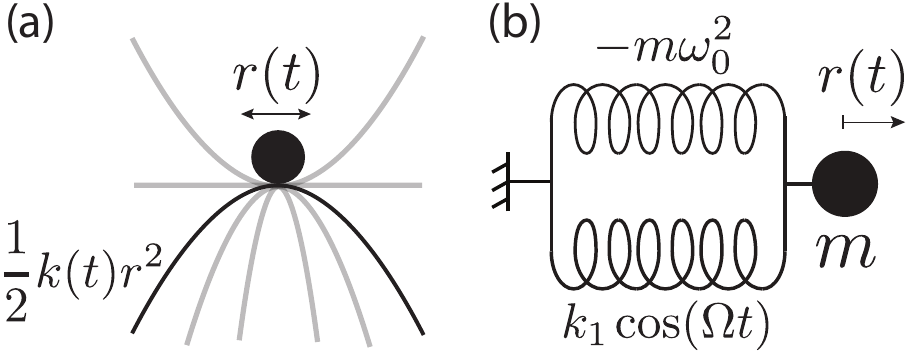}
\caption{The periodically time-varying oscillator under study. (a) Mass on a local quadratic potential hill, harmonically modulated over a period $T=2\pi/\Omega$.  (b) Associated linear mass-spring system with a time-varying stiffness $k(t) = -k_0 + k_1\cos(\Omega t)$ and $k_0=m\omega_0^2$.}
\label{Figure1}
\end{center}
\end{figure}

\section{MATERIALS AND METHODS}

\subsection{Equation of motion and Floquet theory}

Figs. \ref{Figure1}(a),(b) show the $1D$ periodically time-varying linear dynamical system under study. A particle of mass $m$ locally moves with a position $r(t)$ on a harmonically modulated, quadratic potential hill, so that its total kinetic plus potential energy reads $E(t) = 1/2m\dot{r}(t)^2 + 1/2k(t)r(t)^2$ where $\dot{r}(t)$ is the velocity of the particle. The linear equation of motion of $m$ associated with $E(t)$ is modeled by the $1$ degree-of-freedom mass-spring system shown in Fig. \ref{Figure1}(b). In this periodically conservative system, the particle experiences a parametric excitation $F(t)=-k(t)r(t)$ that derives from the modulated potential energy and that can be represented by a spring with a $T$-periodic varying stiffness $k(t)=k(t+T)=-k_0+k_1\cos(\Omega t)$. Here, $-k_0$ and $k_1$ are the fundamental and modulated stiffness, respectively; $\Omega=2\pi/T$ is the frequency of the modulated potential when $\omega_0=\sqrt{k_0/m}$ is a natural frequency. The only difference with a classic parametric oscillator is that the fundamental stiffness $-k_0$ is negative: if $k_1=0$ N/m, the system is linearly unstable and the mass exponentially diverges following $r(t) \propto e^{\omega_0 t}$.

According to Newton's second law and Lagrangian mechanics, the dimensionless Initial Value Problem (IVP) governing $r(t)$ reads
\begin{equation}
\frac{d^2r(\tau)}{d \tau^2}-r(\tau)+\alpha\cos(\tau /\delta)r(\tau) = 0
\label{Eq1}
\end{equation}
where $\tau=\omega_0 t$ is the dimensionless time. In this fundamental Mathieu equation \cite{Whittaker1996,Magnus2013}, the two relevant parameters are the frequency ratio $\delta=\omega_0/\Omega$ and the stiffness ratio $\alpha=k_1/k_0$. For modal and stability analysis of the linear differential Eq.(\ref{Eq1}), one can use Floquet theory \cite{Floquet1879,Whittaker1996} to express $r(\tau)$ as a linear combination of two almost-periodic vibrational modes 
\begin{equation}
r(\tau) = c_1\Psi(\tau)e^{\sigma\tau} + c_2\Psi^{\ast}(\tau)e^{-\sigma\tau}
\label{Eq1bis}
\end{equation}
\noindent
where $c_1$ and $c_2$ are constants determined upon initial position $r(0)$ and velocity $\dot{r}(0)$. Replacing the Floquet form $\Psi(\tau)e^{\sigma\tau}$ in Eq.(\ref{Eq1}) leads to an eigenvalue problem that can be numerically solved for each set of parameters $(\alpha,\delta)$ \cite{Moore2005,Lazarus2010b,Lazarus2017}. The computed eigenfunction $\Psi(\tau)$ and its complex conjugate $\Psi^{\ast}(\tau)$ are periodic with a dimensionless period $\bar{T}=2\pi\delta$. The complex eigenvalue $\sigma$ is called the Floquet exponent. Because of the form of the solution $r(\tau)$ and the absence of damping term in Eq.(\ref{Eq1}), only two qualitative stability behavior can be observed in the $(\alpha,\delta)$ space. Either $\Re(\sigma) = 0$ and $-1/2\delta < \Im(\sigma) \leq 1/2\delta$, which means the particle is neutrally stable and $r(\tau)$ is an almost-periodic oscillation about the stationary state. Or $\Re(\sigma) > 0$ and $\Im(\sigma)=0$ or $\Im(\sigma)=1/2\delta$ and $r(\tau)$ is respectively a $\bar{T}$ or $\bar{2T}$-periodic motion that exponentially oscillates toward infinity with a growth rate $\Re(\sigma)$ meaning the perturbed periodic state is linearly unstable.

\begin{figure}[!t]
\begin{center}
\includegraphics[width=0.8\columnwidth]{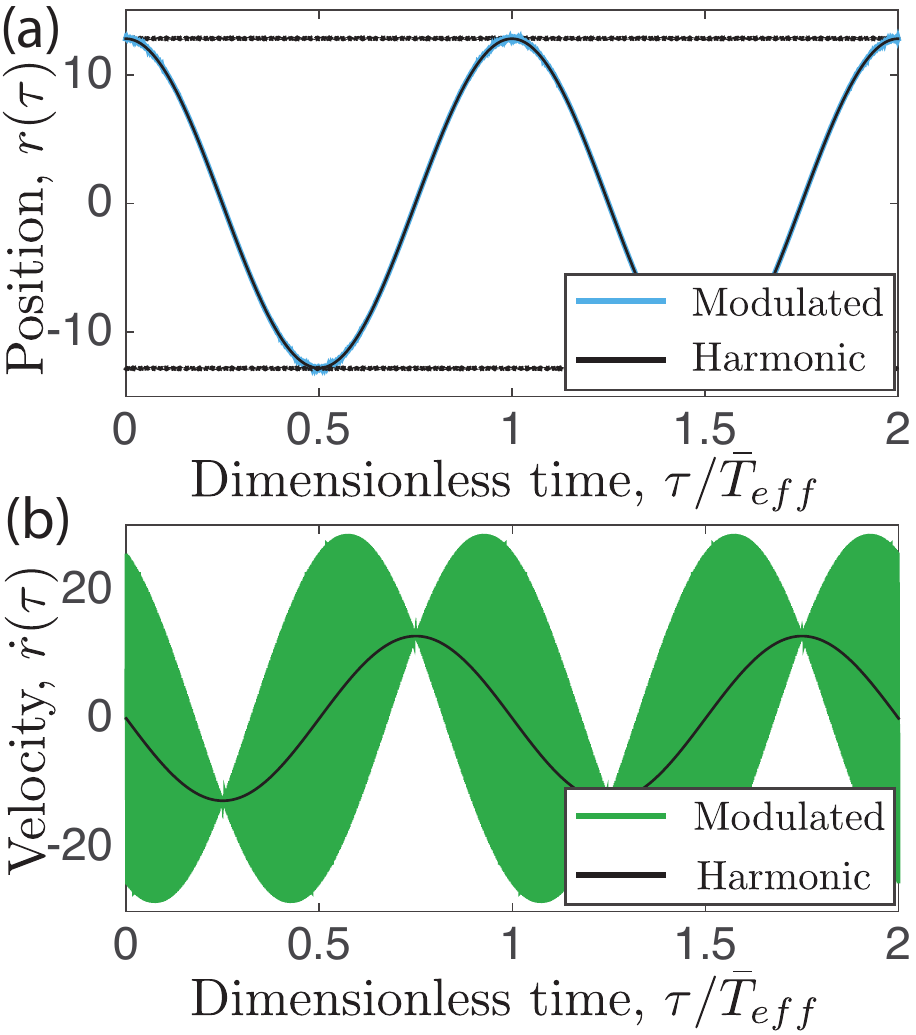}
\caption{Stable vibrational mode in the Kapitza regime for $\delta=1\times 10^{-3}$ and $\alpha = 2000$. (a) Blue line shows the numerical computation of position, $r(\tau)=\Psi(\tau)e^{\sigma \tau}$, as a function of dimensionless time $\tau/\bar{T}_{eff}$ with $\bar{T}_{eff}=2\pi/\omega_{eff}$. The eigenfunction $\Psi(\tau)$ is normalized so that $\int_{\bar{T}}\left|\Psi(\tau)\right|^2d\tau=1$. The moduli $-|\Psi(\tau)|$ and $|\Psi(\tau)|$ are shown in dotted lines and the effective classic harmonic oscillator approximation, $r(\tau)=\Psi(0)\cos(\omega_{eff}\tau)$, is displayed in black line. (b) Green line shows the computed velocity $\dot{r}(\tau)$ when black line represents the effective approximation $\dot{r}(\tau)=-\Psi(0)\omega_{eff}\sin(\omega_{eff}\tau)$. The eigenfunctions $-|\dot{\Psi}(\tau)|$ and $|\dot{\Psi}(\tau)|$ are not shown for a sake of clarity.}
\label{Figure2}
\end{center}
\end{figure}

\subsection{Effective harmonic oscillator in the Kapitza limit $\delta << 1$}

The oscillating system represented by the governing equation Eq.(\ref{Eq1}) has been well studied in the asymptotic limit where the modulation period is much smaller than the collapsing time, i.e. for $\delta << 1$ \cite{Magnus2013}. In this regime, first understood by Kapitza thanks to averaging techniques when studying the inverted pendulum whose pivot point is vertically vibrated \cite{Stephenson1908,Kapitsa1951}, it is possible to stabilize the naturally collapsing mass if $(\alpha \delta)^2/2 > 1$. In this particular stable regime where $\Re(\sigma)=0$, the harmonically modulated oscillator of Fig. \ref{Figure1} behaves like an effective classic harmonic oscillator whose effective natural frequency is given by
\begin{equation}
\omega_{eff} = \sqrt{\frac{(\alpha \delta)^2}{2} - 1}.
\label{Eq1ter}
\end{equation}
\noindent
Fig. \ref{Figure2} shows the typical Floquet form or oscillating mode of Eq.(\ref{Eq1}), $r(\tau) = \Psi(\tau)e^{\sigma\tau}$, for a duo of modulation parameters in the Kapitza's stabilization regime (for practical purposes, we chose $\alpha = k_1/k_0 = 2000$ and $\delta = \omega_0/\Omega = 1 \times 10^{-3}$). The mode is normalized so that $N=\int_{\bar{T}}\Psi^{\ast}(\tau)\Psi(\tau)d\tau = \int_{\bar{T}}\left|\Psi(\tau)\right|^2d\tau=1$ and only the physical response, i.e. the real part of $\Psi(\tau)e^{\sigma\tau}$, is shown. Note that in the neutrally stable regime, $(\alpha \delta)^2/2 > 1$, where $\sigma$ is a pure imaginary number, the second vibrational mode $r_2(\tau)=\Psi^{\ast}(\tau)e^{-\sigma\tau}$ of Eq.(\ref{Eq1bis}) is simply conjugate or out-of-phase with the displayed one. Fig. \ref{Figure2}(a) displays in blue line the position over two effective periods $\bar{T}_{eff}=2\pi/\omega_{eff}$ which corresponds to $2/\delta=2000$ fast modulation periods. We see that because of the averaging of the modulation, one can model the position of the almost-periodically vibrating mass by the one of the effective classic harmonic oscillator, $r(\tau)=\Psi(0)\cos(\omega_{eff}\tau)$, shown in black line. Also, the dotted lines that represent the $\bar{T}$-periodic envelopes of the Floquet form, $-|\Psi(\tau)|$ and $|\Psi(\tau)|$, tend to the constant envelopes, $-|\Psi(0)|$ and $|\Psi(0)|$ of the position of the effective classic harmonic oscillator. Fig. \ref{Figure2}(b) shows in green line the associated velocity of the particle. The velocity of the effective classic harmonic oscillator, $\dot{r}(\tau)=-\Psi(0)\omega_{eff}\sin(\omega_{eff}\tau)$, given in black line, only models the averaged velocity, that is actually strongly oscillating. In the next Section, we will focus on the qualitative behavior of the modulated oscillator of Fig. \ref{Figure1} when $\delta > 1$ that has been overlooked.

\begin{figure}[!t]
\begin{center}
\includegraphics[width=0.815\columnwidth]{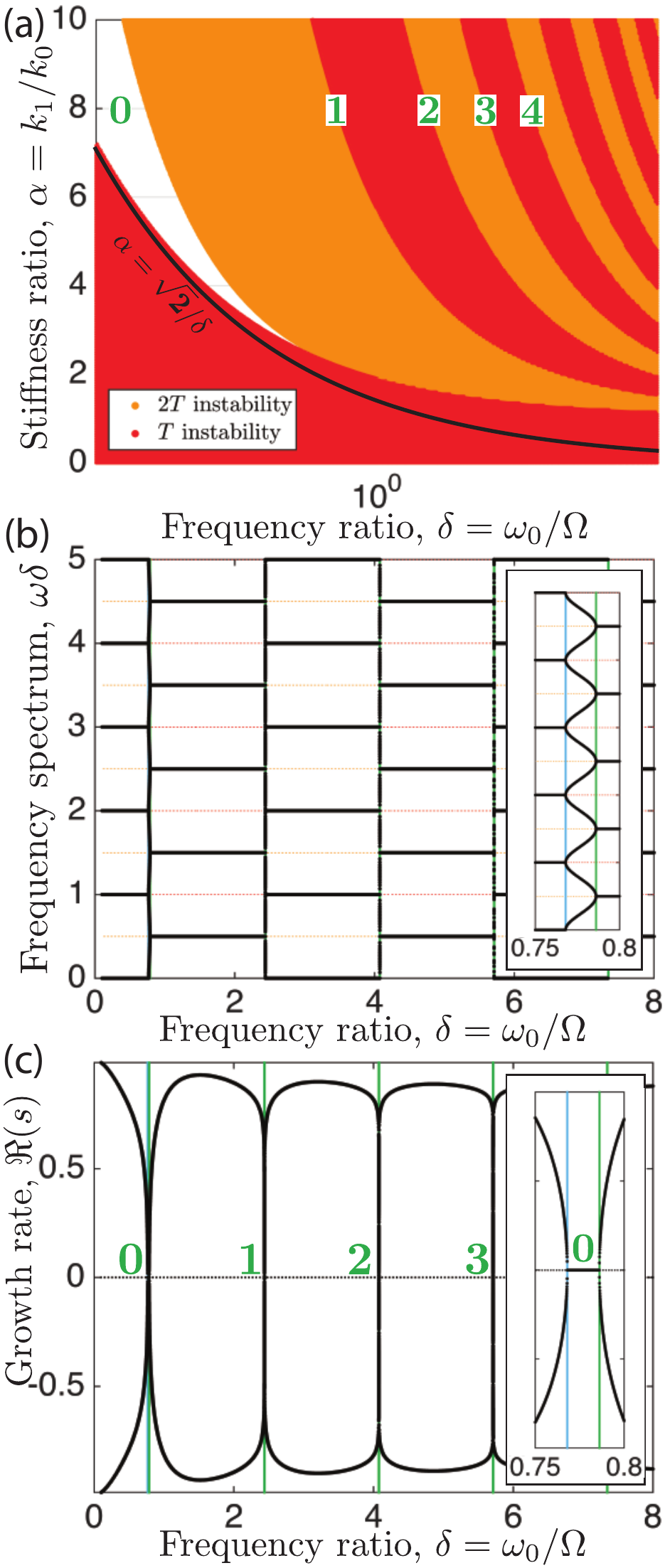}
\caption{Numerical stability analysis of the harmonically modulated collapsing mass.  (a) Linear stability chart for $0 \leq \alpha \leq 10$ and $0.05 \leq \delta \leq 14.3$. Red and yellow regions show $\bar{T}$ and $2\bar{T}$-periodic unstable solutions, respectively. Between those regions are tiny zones, denoted by $l=0,1,2,\ldots$ in ascending order of $\delta$, where $r(\tau)$ is a neutrally stable almost-periodic solution. The black line represents the Kapitza limit $\alpha = \sqrt{2}/\delta$. (b) Location of the frequency spectrum $\omega$ of the solution $r(\tau)$ as a function of $\delta$ for $\alpha = 2.25$. Green and blue vertical lines show the neutrally stable $\bar{T}$ and $2\bar{T}$-periodic solutions, respectively. Inset: Zoom on the first stability region $l=0$ where the spectrum of the two Floquet forms are unlocked. (c) Evolution of the growth rate of the two vibrational modes $\Psi(\tau)e^{\sigma\tau}$ and $\Psi^{\ast}(\tau)e^{-\sigma\tau}$ as a function of $\delta$ for $\alpha = 2.25$. Inset: Zoom on the first stability region $l=0$ where the growth rate of both modes is $\Re(\sigma)=0$.}
\label{Figure3}
\end{center}
\end{figure}

\section{MODAL AND STABILITY ANALYSIS ($\delta > 1$)}

\subsection{Quantization of the stable vibrational modes in the modulation parameter space $(\alpha,\delta)$}

To get a better physical understanding of the oscillator of Fig. \ref{Figure1}, we perform a numerical stability analysis of Eq.(\ref{Eq1}) by analyzing the growth rate $\sigma$ of the two computed Floquet forms $\Psi(\tau)e^{\sigma\tau}$ and $\Psi^{\ast}(\tau)e^{-\sigma\tau}$ in the modulation parameter space $(\alpha,\delta)$. Notably, according to the superposition property of Eq.(\ref{Eq1bis}), the system is unstable or the position of the mass $r(\tau)$ is diverging with a period $\bar{T}=2\pi\delta$ or $2\bar{T}$ if $\Re(\sigma) \neq 0$ and $\Im(\sigma)=0$ or $\Im(\sigma)=1/2\delta$, respectively. Fig. \ref{Figure3}(a) shows the linear stability chart of the particle for $0 \leq \alpha \leq 10$ and $0.05 \leq \delta \leq 14.3$. Like for a classic Mathieu equation \cite{Whittaker1996,Magnus2013}, there is an alternation of $\bar{T}$-unstable (red dots in Fig. \ref{Figure3}(a)) and $2\bar{T}$-unstable regions (yellow dots). If the system is not modulated enough, i.e. for $\alpha < 1$, the mass cannot be stabilized. As explained in the previous section, for $\delta << 1$, a stability region opens whose instability lower limit corresponds to the classic relation of Kapitza, $\alpha = \sqrt{2}/\delta$, shown in black line in Fig. \ref{Figure3}(a). But for $\delta > 1$, the stabilization of the mass is still theoretically possible albeit in a discrete fashion in the $(\alpha,\delta)$ modulation space. According to Floquet theory, the $\bar{T}$ and $2\bar{T}$ instability regions cannot merge in the $(\alpha,\delta)$ space so each alternation of colors indicates a tiny stability domain that we denote $l=0,1,2,\ldots$ as explained in Fig. \ref{Figure3}(a) \cite{Movies}. Thus, several regions of stability form independent ``branches'' whose width drastically decrease as $\delta$ increases (we barely reach inside the stability regions above $\delta>6$ because of machine epsilon of our computational software).

Figs. \ref{Figure3}(b) and (c) give a physical insight in the aforementioned stability mechanism since they show the evolution of the spectrum of the two Floquet forms $r(\tau) = \Psi(\tau)e^{\sigma\tau}$ and $r(\tau) = \Psi^{\ast}(\tau)e^{-\sigma\tau}$ as a function of $\delta$ for $\alpha = 2.25$. According to Floquet theory, the eigenfunction $\Psi(\tau)$ is $\bar{T}$-periodic with a fundamental frequency $1/\delta$. Therefore, the spectrum of the two Floquet forms read $\sigma + \sum_h h/\delta$ and $-\sigma +\sum_h h/\delta$, where $h$ is an integer. By the superposition principle of Eq.(\ref{Eq1bis}), the solution $r(\tau)$ contains the sum of both spectrum in it. Fig. \ref{Figure3}(b) shows the location of the frequency spectrum $\omega$ of the two vibrational modes $\Im(\sigma + \sum_h h/\delta)$ and $\Im(-\sigma + \sum_h h/\delta)$ as a function of $\delta$ for $\alpha = 2.25$ (only the positive part of the spectrum is shown as the latter is symmetric with respect to the $x$-axis). For most $\delta$, the frequency spectra of the two modes are locked, alternatively in $0 + \sum_h h/\delta$ ($\bar{T}$-periodic spectrum) or in $1/2\delta + \sum_h h/\delta$ ($2\bar{T}$-periodic spectrum). As shown in Fig. \ref{Figure3}(c), this lock-in is associated with a growth rate $\Re(\sigma)$ that departs from $0$ so that one of the periodically oscillating Floquet form is exponentially diverging ($\Re(\sigma) > 0$) and one is damped ($\Re(\sigma) < 0$). Between those locked unstable regions, tiny zones exist where the two Floquet modes are unlocked and where the associated growth rate $\Re(\sigma)$ is zero for both modes, a situation that corresponds to neutrally stable almost periodic solutions $r(\tau)$. The insets in Figs. \ref{Figure3}(b) and (c) display a zoom on the first stability region $l=0$. In stability regions, the fundamental frequency $\Im(\sigma)$ varies continuously from $0$ to $1/2\delta$ as $\delta$ varies. Note that the green and blue vertical lines are the limits of the stability zones and correspond to neutrally stable $\bar{T}$ and $2\bar{T}$-periodic solutions with $\Re(\sigma)=0$.

\begin{figure}[!t]
\begin{center}
\includegraphics[width=0.85\columnwidth]{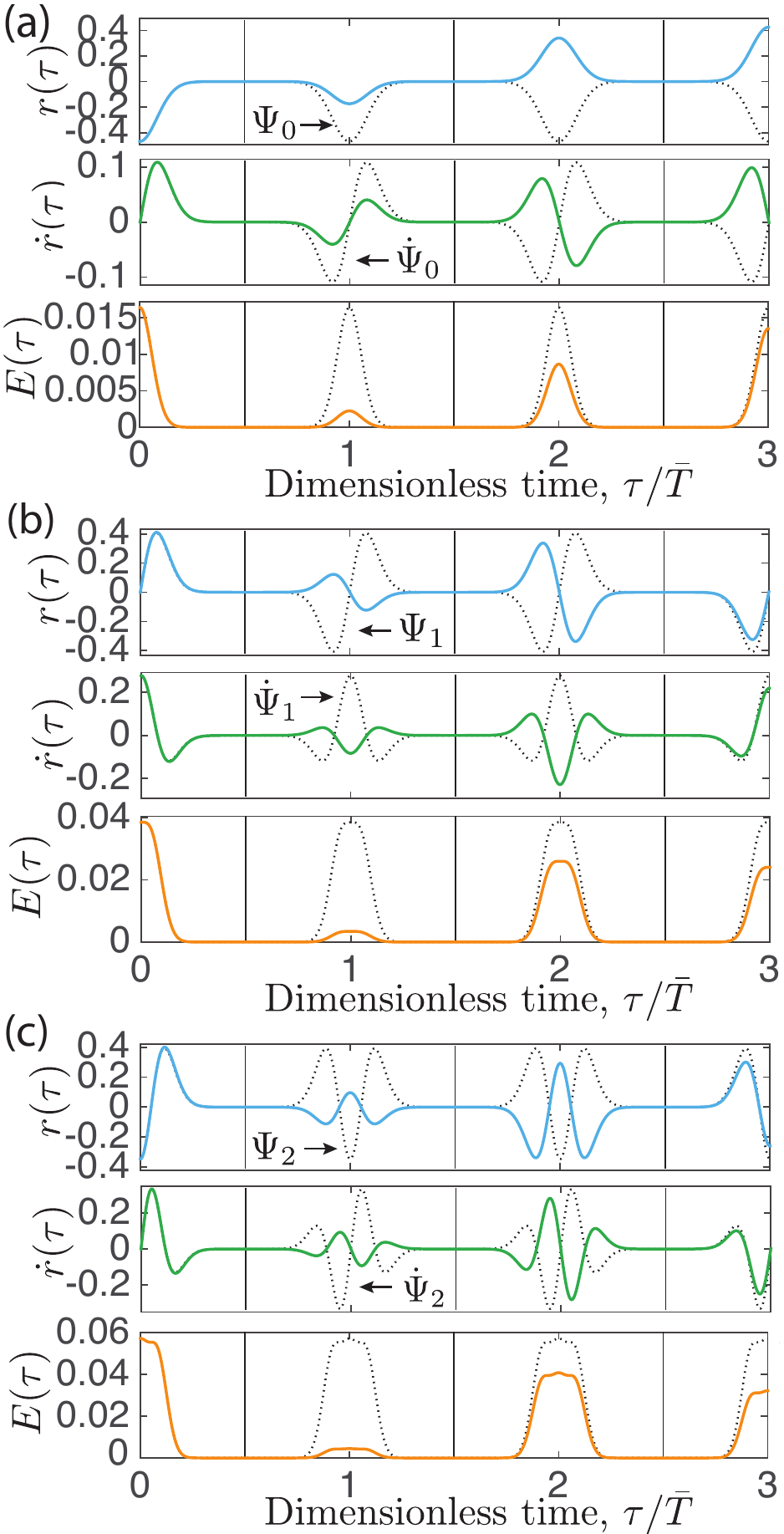}
\caption{Time evolution position $r(\tau)$, velocity $\dot{r}(\tau)$ and energy $E(r(\tau))$ of neutrally stable vibrational modes $\Psi(\tau)e^{i\Im(\sigma) \tau}$ with $N=\int_{\bar{T}}|\Psi(\tau)|^2d\tau =1$. Periodic eigenfunctions $\Psi(\tau)$, $\dot{Psi}(\tau)$ and $E(\Psi(\tau))$ are shown in dotted lines. Periodic cells are delimited by thin vertical lines. (a) Mode in the $l=0$ stability regions for $\delta = 5$ and $\alpha = 1.14905152323442$. (b) Mode in the $l=1$ stability regions for $\delta = 5$ and $\alpha = 1.5081802859941$. (c) Mode in the $l=2$ stability regions for $\delta = 5$ and $\alpha = 1.9549540063305$.}
\label{Figure4}
\end{center}
\end{figure}

We now focus on the neutrally stable motion $(r(\tau),\dot{r}(\tau))$ of the mass to characterize the allowed vibrational modes of the modulated oscillator under study in the regime $\delta > 1$. Figs. \ref{Figure4}(a), (b) and (c) display three typical examples of such vibrational modes over three periods $\bar{T}$ for modulation parameters in the stability regions $l=0$, $1$ and $2$, respectively \cite{Movies}. In each panel, we show the position $r(\tau)=\Psi(\tau)e^{i\Im(\sigma) \tau}$, velocity $\dot{r}(\tau)$ and mechanical energy $E(r(\tau))= 1/2\dot{r}(\tau)^2 + 1/2(-1+\alpha\cos(\tau/\delta))r(\tau)^2$ of the computed vibrational mode. The modes are normalized so that $N=\int_{\bar{T}}\Psi^{\ast}(\tau)\Psi(\tau)d\tau =1$. Full lines represent the computed almost-periodic motion when dotted lines show the periodic eigenfunctions $\Psi(\tau)$, $\dot{\Psi}(\tau)$ and $E(\Psi(\tau))$. The stable almost-periodic vibrations of the mass described in Fig. \ref{Figure4} can be decomposed in a successive repetition of similar motions that are scaled copies of their $\bar{T}$-periodic eigenfunctions (the scaling factors can take all the values between $-1$ and $1$ and will be discussed in next section). The $\bar{T}$-periodic eigenfunctions can themselves be decomposed in three parts: (1) when the periodic stiffness $\bar{k}(\tau) = (-1+\cos(\tau/\delta))$ is negative, the mass exponentially diverges, (2) when $\bar{k}(\tau)>0$ the mass oscillates, (3) when $\bar{k}(\tau)<0$ the mass exponentially converges to a state very almost identical to the previous period \cite{Movies}. The qualitative difference between vibrational motions with ascending order of $l$ from Fig. \ref{Figure4}(a) to (c) is that in the $l^{th}$ stability region, the mass is able to do $(l+1)$ oscillations during the time $\bar{k}(\tau)$ is positive. Like in a classical oscillator, the velocity is zero when the position is at a local maximum. Finally, the mechanical energy $E(r(\tau))$ characterizes the binary qualitative behavior of the studied oscillator: over one period $\bar{T}$, it seems null when $\bar{k}(\tau)<0$ and positive when $\bar{k}(\tau)>0$.
\\

\subsection{Analogy with the uncertainty relations of the quantum harmonic oscillator}

So far, we have been qualitative on the description of the $1D$ harmonically modulated linear oscillator of Fig. \ref{Figure1} for $\delta > 1$ and have shown that the allowed motions get quantized in particular vibrational modes in the modulation parameter space. Now, we quantitatively characterize the modes of Fig. \ref{Figure4} thanks to an analogy with the statistical interpretation of the quantum harmonic oscillator \cite{Messiah1961} that is illustrated in appendix A. Assume we have a measuring device that detects the presence of the particle but only above a triggering amplitude of motion, $r(\tau)=r_0$. We define $P(\tau)d\tau$ as the probability to find the particle in the infinitesimal time interval $[\tau,\tau+d\tau]$. By analogy with quantum mechanics, we determine the probability density $P(\tau)$ by the relation
\begin{equation}
P(\tau) = \Psi^{\ast}(\tau)\Psi(\tau)=\left|\Psi(\tau)\right|^2.
\label{Eqproba}
\end{equation}

\begin{figure}[!t]
\begin{center}
\includegraphics[width=0.77\columnwidth]{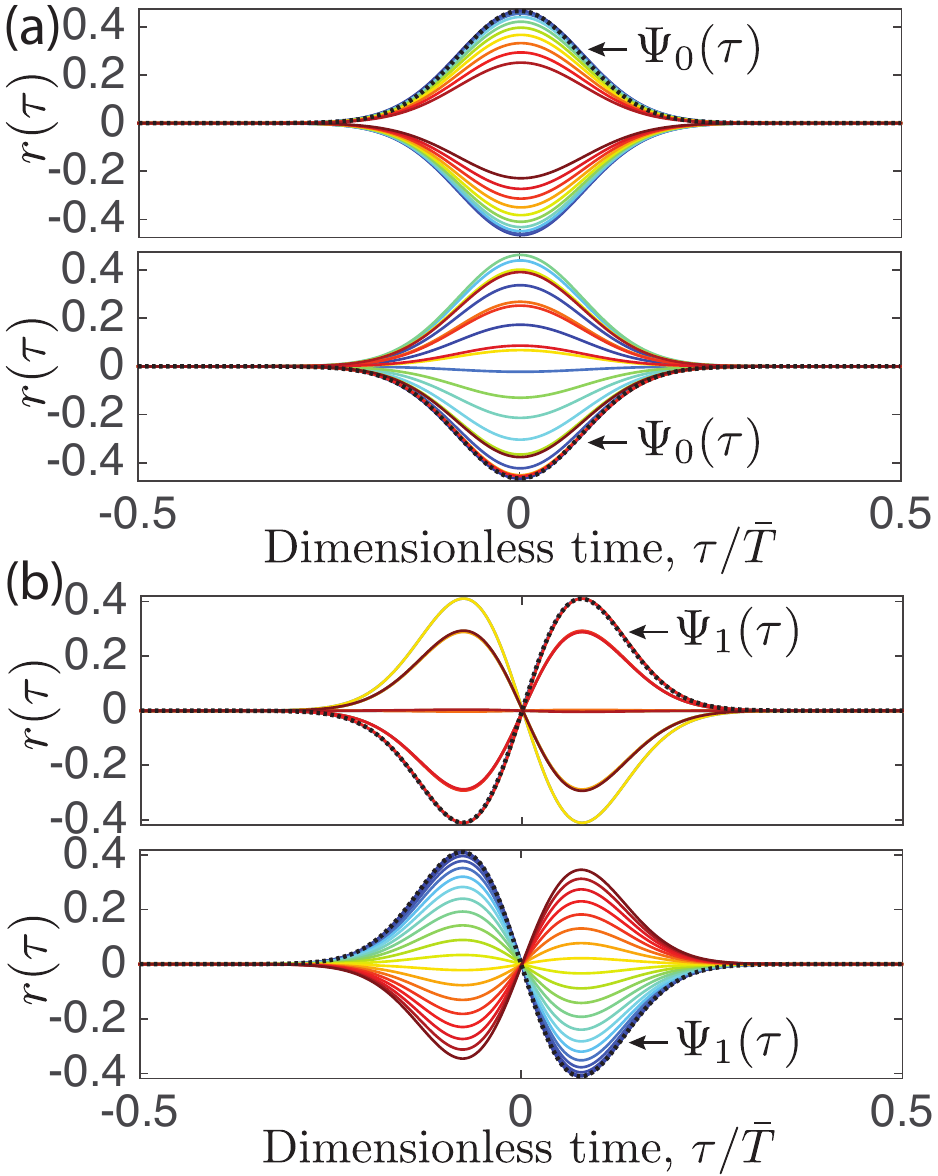}
\caption{Sensitivity analysis of the stable vibrational modes $r(\tau)=\Psi(\tau)e^{i\Im(\sigma)\tau}$ for $\delta = 5$. (a) For a mode in the $l=0$ stability region, the $20$ first consecutive periods of $r(\tau)$, chronologically ordered from cold blue to hot red, are superposed on the primitive time cell $-\bar{T}/2 \leq \tau < \bar{T}/2$. Dotted lines represent the periodic eigenfunction $\Psi_0(\tau)$. Top and bottom panel have slightly different $\delta$ corresponding to $\Im(\sigma) \approx 0.98\times 1/2\delta$ and $\Im(\sigma) \approx 0.62\times 1/2\delta$, respectively. (b) Same as (a) but for a mode in the stable zone $l=1$. Top and bottom panel have slightly different $\delta$ corresponding to $\Im(\sigma) \approx 0.25\times 1/2\delta$ and $\Im(\sigma) \approx 0.04\times 1/2\delta$, respectively. }
\label{Figure5}
\end{center}
\end{figure}

\noindent
Since $r(\tau)$ can be decomposed in successive ``seemingly random'' scaled copies, between $-1$ and $1$, of the $\bar{T}$-periodic eigenfunction $\Psi(\tau)$ as illustrated in Fig. \ref{Figure4} and \ref{Figure5}, the definition Eq.(\ref{Eqproba}) conveys the idea that the probability of measuring the presence of the particle at time $\tau$ is stronger when the relative intensity of $\Psi(\tau)$ is stronger (we will gradually clarify this notion of ``apparent randomness'' in next section). Moreover, in account of the $\bar{T}$-periodicity of $\Psi(\tau)$, $P(\tau)=P(\tau+\bar{T})$ and one can reduce the statistics to a single representative period. Assume also a measurement has been made, the total probability to detect the particle over the representative period $\bar{T}$ would then be $N=\int_{\bar{T}}P(\tau)d\tau =1$ which is the normalization condition that we have chosen in Fig. \ref{Figure4}. Knowing the distribution $P(\tau)$ over the period $\bar{T}$, it is possible to define the average value a continuous function of interest, $F(\tau)$, would take after independently measuring it a very large number of time 
\begin{equation}
\langle F(\tau) \rangle = \int_{\bar{T}}P(\tau)F(\tau)d\tau = \int_{\bar{T}}\Psi^{\ast}(\tau)F(\tau) \Psi(\tau)d\tau.
\label{Eqmean}
\end{equation}
The standard deviation, illustrating the statistical fluctuation of the measures of $F(\tau)$ around the average $\langle F(\tau) \rangle$, would be expressed by
\begin{equation}
\Delta_{F}=\sqrt{\langle F(\tau)^2 \rangle-\langle F(\tau) \rangle^2}.
\label{Eqstandarddeviation}
\end{equation}

By taking $F(\tau)=\tau$ and $F(\tau)=\tau^2$, it is possible to compute $\Delta_{\tau}$, a statistical fluctuation of the measure of $\tau$ around $\langle \tau \rangle$, for the three stable vibrational modes of the stability regions $l=0$, $1$ and $2$ shown in Fig. \ref{Figure4}(a), (b) and (c). The probability to find $\tau$ so that $r(\tau)>r_0$ would not solely depend on the shape of the periodic eigenfunctions $\Psi(\tau)$, it would also depend on the values of the initial conditions $(r(0),\dot{r}(0))$, i.e. on the mechanical energy $E(r(0))$ stored in the linear oscillator (see appendix A). By choosing $F(\tau)=\hat{E}=i\frac{\partial}{\partial \tau}$ which is the energy operator of quantum mechanics for a reduced Planck constant $\hbar=1$, and therefore $F(\tau)=\hat{E}^2=-\frac{\partial^2}{\partial \tau^2}$, one can compute $\Delta_{\hat{E}}$ from Eqs.(\ref{Eqmean})-(\ref{Eqstandarddeviation}). Numerically estimating the product of standard deviations $\Delta_{\tau}\Delta_{\hat{E}}$ for each mode, we find, with a precision of $10^{-4}$, 
\begin{equation}
\Delta_{\tau}\Delta_{\hat{E}}=l+\frac{1}{2},
\label{Equncertainty}
\end{equation}
which is the time-energy uncertainty relations of the quantum harmonic oscillator where $l$ denotes the mode number \cite{Messiah1961}. Note that the computed result in Eq.(\ref{Equncertainty}) is independent on the chosen interval $[\tau,\tau+\bar{T}]$ in Fig. \ref{Figure4}. Repeating the calculation $\Delta_{\tau}\Delta_{\hat{E}}$ for many more computed modes in the first four $l^{th}$ stability regions for $\delta>1$ and $\alpha < 8$, we find the same result given in Eq.(\ref{Equncertainty}) with a maximum error of $10^{-3}$. The relation gets more accurate as $\delta$ increases and the width of the stability regions decreases. The analog of the time-energy uncertainty relation Eq.(\ref{Equncertainty}) is therefore a quantitative property verified by the stable vibrational modes of our oscillator in a certain $(\alpha,\delta)$ limit. In the next section, we strive to get a deeper physical and mathematical insight in the description of the periodic eigenfunctions on the $l^{th}$ stability region, $\Psi_l(\tau)$, and their relation with the actual position $r(\tau)$.

\section{REDUCTION OF THE DYNAMICS TO $\Psi(\tau)$}

\subsection{\label{reduc1} Reduction to the neutrally stable periodic solutions $r_l(\tau)=\Psi_l(\tau)$}

As already mentioned and illustrated in Fig. \ref{Figure4}, the almost-periodic neutrally stable modes of the $l^{th}$ stability region, $r_l(\tau)=\Psi_l(\tau)e^{i\Im(\sigma)\tau}$, can be decomposed in a succession of cycles that are scaled copies of the periodic eigenfunctions $\Psi_l(\tau)=\Psi_l(\tau+\bar{T})$. To highlight this time-translational property, we superpose, on the primitive time cell $-\bar{T}/2 \leq \tau < \bar{T}/2$, the $20$ first consecutive periods of the position $r(\tau)$ of some computed vibrational modes. The result is shown with $\delta=5$ and $\alpha = 1.14905152323442$ for modes in the $l=0$ stability region in Fig. \ref{Figure5}(a) where the color of the $20$ lines from cold blue to hot red indicates an increase of period. The dotted line represents the periodic eigenfunction $\Psi_0(\tau)$. Around $\delta=5$, the width $\epsilon$ of the stability region has already decreased to $\epsilon \approx 10^{-13}$ in the $(\alpha,\delta)$ parameter space. As a consequence, the eigenvalue or fundamental frequency $\Im(\sigma)$ varies extremely rapidly in the primitive spectral cell $0 \leq \Im(\sigma) \leq 1/2\delta$ as a function of $(\alpha,\delta)$ (see insets of Fig. \ref{Figure3}(b) and (c) to see the evolution of $\sigma$ in a stability region); unlike the periodic eigenfunction $\Psi(\tau)$ that varies in the order of a stability region to another. As a consequence, the position of the stable vibrational mode, $r(\tau)=\Psi(\tau)e^{i\Im(\sigma)\tau}$ is sensitive to the modulation parameter $(\alpha,\delta)$ but not its periodic eigenfunction (the same is true for $\dot{r}(\tau)$ and $\dot{\Psi}(\tau)$ or $E(r(\tau))$ and $E(\Psi(\tau))$). This sensitivity property is highlighted in the top and bottom panels of Fig. \ref{Figure5}(a) that show vibrational modes in the $l=0$ stability region for $\delta = 5 \pm \epsilon/10$, corresponding to $\Im(\sigma) \approx 0.98\times 1/2\delta$ and $\Im(\sigma) \approx 0.62\times 1/2\delta$, respectively. From one panel to the other, $\Psi_0(\tau)$ as well as the various curves of $r(\tau)$ have similar shapes, the only change lies in the chronological order of the scaling factor with respect to $\Psi(\tau)$. The same numerical observations on sensitivity can be made for vibrational modes in higher regions of stability as illustrated for $l=1$ in Fig. \ref{Figure5}(b) ($\delta = 5$ and $\alpha = 1.9549540063305$), where the top and bottom panel correspond to a fundamental frequency $\Im(\sigma) \approx 0.25\times 1/2\delta$ and $\Im(\sigma) \approx 0.04\times 1/2\delta$. Note that a similar reasoning can be made in the frequency domain as shown in Appendix B.

\begin{figure}[!t]
\begin{center}
\includegraphics[width=0.77\columnwidth]{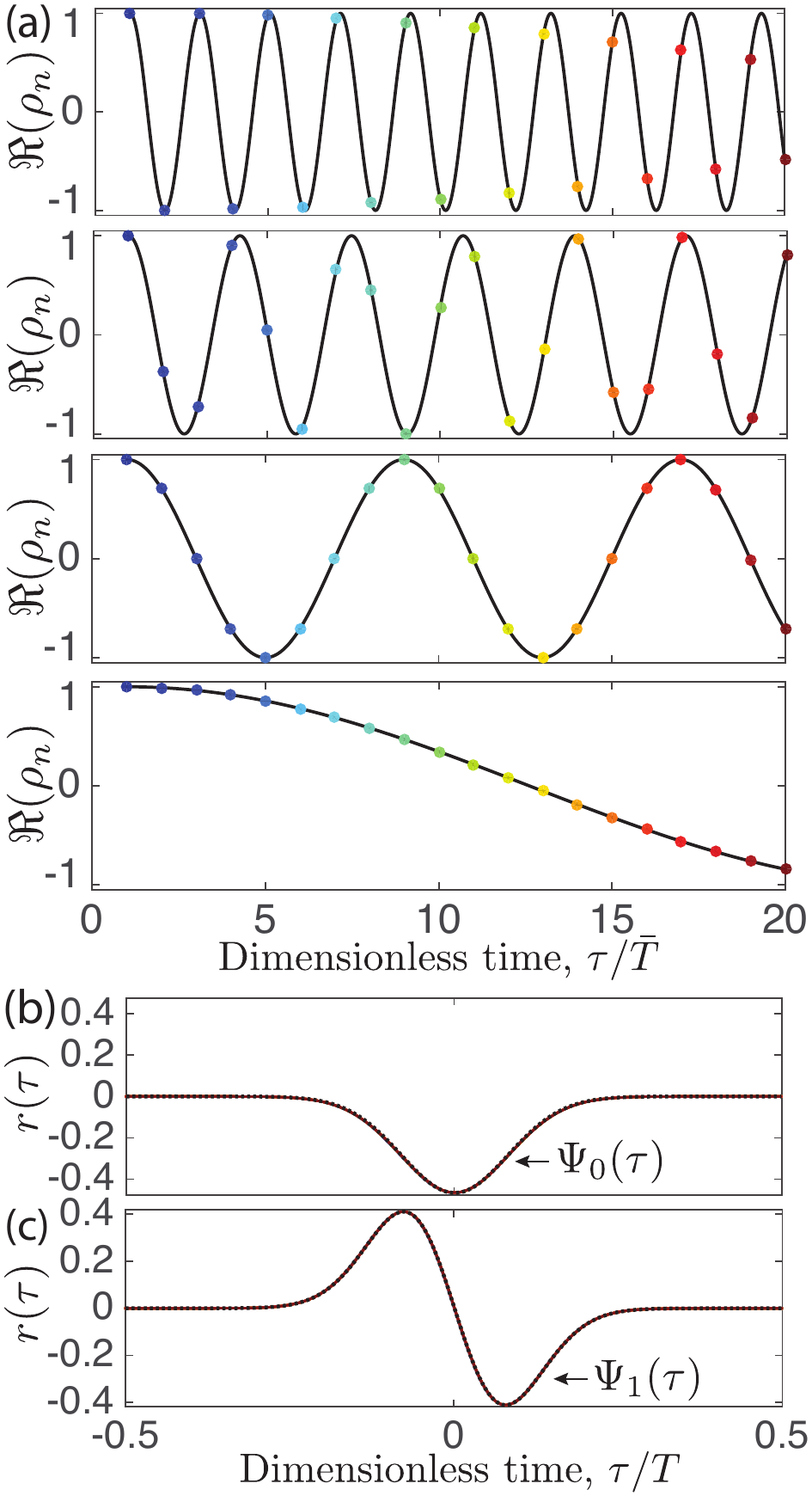}
\caption{Periodic mapping and reduction of the neutrally stable vibrational modes to their periodic eigenfunctions. (a) Discrete evolution of the $n=20$ first scaling factors $\Re(\rho_n)$ as a function of number of periods. Black lines show the functions $\cos(\Im(\sigma)(\tau-\bar{T}))$ and color dots represent the Floquet multipliers $\Re(\rho_n)= \cos(\Im(\sigma)(n\bar{T}-\bar{T}))$. From top to bottom we have $\Im(\sigma) \approx 0.98\times 1/2\delta$, $\Im(\sigma) \approx 0.62\times 1/2\delta$, $\Im(\sigma) \approx 0.25\times 1/2\delta$ and $\Im(\sigma) \approx 0.04\times 1/2\delta$ with $\delta = 5$. (b) Reduction of the positions $r(n\bar{T}-\bar{T},n\bar{T})$ of Fig. \ref{Figure5}(a) on the eigenfunction $\Psi_0(\tau)$ by scaling by $\Re(\rho_n)$. (c) Same as (b) but for the eigenfunction $\Psi_1(\tau)$ of Fig. \ref{Figure5}(b).}
\label{Figure6}
\end{center}
\end{figure}

To go further, one needs to specify the scaling factors that relates the position $r(\tau)=\Psi(\tau)e^{i\Im(\sigma)\tau}$  to the eigenfunction $\Psi(\tau)$ on each successive periods. When dealing with classic Floquet forms \cite{Lazarus2017}, the almost-periodic solution, $r(\tau)$, theoretically verifies the periodic mapping 
\begin{equation}
r(n\bar{T}) = r(0) \times \Re(\rho_n) = \Psi(0) \times \Re(\rho_n),
\label{EqFloquetmulti}
\end{equation}
where $n$ is a positive integer, $\rho_n=e^{i\Im(\sigma)n\bar{T}}$ is the so-called Floquet multiplier so that $\Re(\rho_n)=\cos(\Im(\sigma)n\bar{T})$ and we recall $\bar{T}=2\pi\delta$ is the modulation period. What we observe in the numerical results illustrated in Fig. \ref{Figure5} is stronger than Eq.(\ref{EqFloquetmulti}) since it is the full function $r(n\bar{T}-\bar{T},n\bar{T})$, i.e. the function $r(\tau)$ in the range $[n\bar{T}-\bar{T},n\bar{T}]$, that is related to the $\bar{T}$-periodic function $\Psi(\tau)$. The mathematical reason for this is beyond the scope of this article but from numerical observations of $\Psi(\tau)$, we are able to generalize the quasi-periodicity property of Eq.(\ref{EqFloquetmulti}) from a single time value $\tau$ to a full period $[\tau,\tau+\bar{T}]$ so that
\begin{equation}
r(n\bar{T}-\bar{T},n\bar{T}) = \Psi(\tau) \times \cos(\Im(\sigma)(n\bar{T}-\bar{T})).
\label{EqFloquetmulti2}
\end{equation}
In Fig. \ref{Figure6}(a), we show the evolution of $\cos(\Im(\sigma)(\tau-\bar{T}))$ as well as the scaling factors $\Re(\rho_n)= \cos(\Im(\sigma)(n\bar{T}-\bar{T}))$ represented by color dots. From top to bottom, we varied, for $\delta=5$, $\Im(\sigma)$ in the primitive spectral cell $0 \leq \Im(\sigma) \leq 1/2\delta$ by taking $\Im(\sigma) \approx 0.98\times 1/2\delta$, $\Im(\sigma) \approx 0.62\times 1/2\delta$, $\Im(\sigma) \approx 0.25\times 1/2\delta$ and $\Im(\sigma) \approx 0.04\times 1/2\delta$, respectively. Those values of $\Im(\sigma)$ are the ones we obtained in Fig. \ref{Figure5} from top to bottom. In Fig. \ref{Figure6}(b) and (c), we divide each position function $r(n\bar{T}-\bar{T},n\bar{T})$ of Fig. \ref{Figure5}(a) and (b) by the corresponding Floquet multipliers $\Re(\rho_n)$ of Fig. \ref{Figure6}(a): the $20$ first periods of the various computed $r_l(\tau)$ all collapse on their respective eigenfunctions $\Psi_l(\tau)$. If we increase the number of periods to a very large number, the scaling factors $\Re(\rho_n)$ take a very large number of different values between $-1$ and $1$ and all the $r(n\bar{T}-\bar{T},n\bar{T})$ would collapse on $\Psi(\tau)$. Therefore, the knowledge of $\Psi(\tau)$ and $\Im(\sigma)$ completely determines $r(\tau)$ by Eq.(\ref{EqFloquetmulti2}).

The property of Eq.(\ref{EqFloquetmulti2}), altogether with the aforementioned sensitivity property of thin regions of stability when $\delta>1$, offers a reduction opportunity. The $l^{th}$ region of stability in the $(\alpha,\delta)$ space as shown in Fig. \ref{Figure3}, can be reduced to a single branch since we know $\Psi_l(\tau)$ as well as the limits of the primitive spectral cell $0 \leq \Im(\sigma) \leq 1/2\delta$ are insensitive to the width $\epsilon$ of the regions. The actual vibrational mode $r_l(\tau)$ could then be retrieved by Eq.(\ref{EqFloquetmulti2}). However, here emerges an analogy with the wave-particle duality of quantum mechanics. One could choose to treat the oscillator of Fig. \ref{Figure1} as a classical dynamical system, i.e. for given initial conditions and given modulation parameters $(\alpha,\delta)$, we compute the neutrally stable time evolution of the particle. But it becomes complicated to follow the various stable modes due to machine precision since as $\delta$ increases, the width $\epsilon$ of the stability regions rapidly decreases. Or, we could choose to characterize the stable modes of vibration by their ``wavy'' eigenfunctions $\Psi_l(\tau)$, i.e. to reduce the $l^{th}$ stability region into a branch with no width and infer the position $r_l(\tau)$ from Eq.(\ref{EqFloquetmulti2}). But, doing so, we would loss information about the causality of the original dynamical system. We would know all the possible motions $r_l(\tau)$ because we know from Floquet theory that $\Im(\sigma)$ is in $0 \leq \Im(\sigma) \leq 1/2\delta$, but we would discard which $\Im(\sigma)$ it is, i.e. which successive scaling factors $\Re(\rho_n)=\cos(\Im(\sigma)(n\bar{T}-\bar{T}))$ the vibrational motion $r_l(\tau)$ has picked. A possible approach would be to represent our ignorance by randomly varying $\Re(\rho_n)$ between $-1$ and $1$, a plausible reality for $\delta >> 1$ where the stability regions becomes so thin in $(\alpha,\delta)$ that the precision of a computing device on $\Im(\sigma)$ could fluctuate in the spectral cell $0 \leq \Im(\sigma) \leq 1/2\delta$ (the statistical approach of the previous section could then be acceptable). In next subsection, we show that when reducing the stability regions on the branches of purely periodic solutions, i.e. $\sigma = 0$ and $r_l(\tau)=\Psi_l(\tau)e^{\sigma\tau}=\Psi_l(\tau)$, one can get an analytical expression for $\Psi_l(\tau)$.

\begin{figure}[!b]
\begin{center}
\includegraphics[width=0.82\columnwidth]{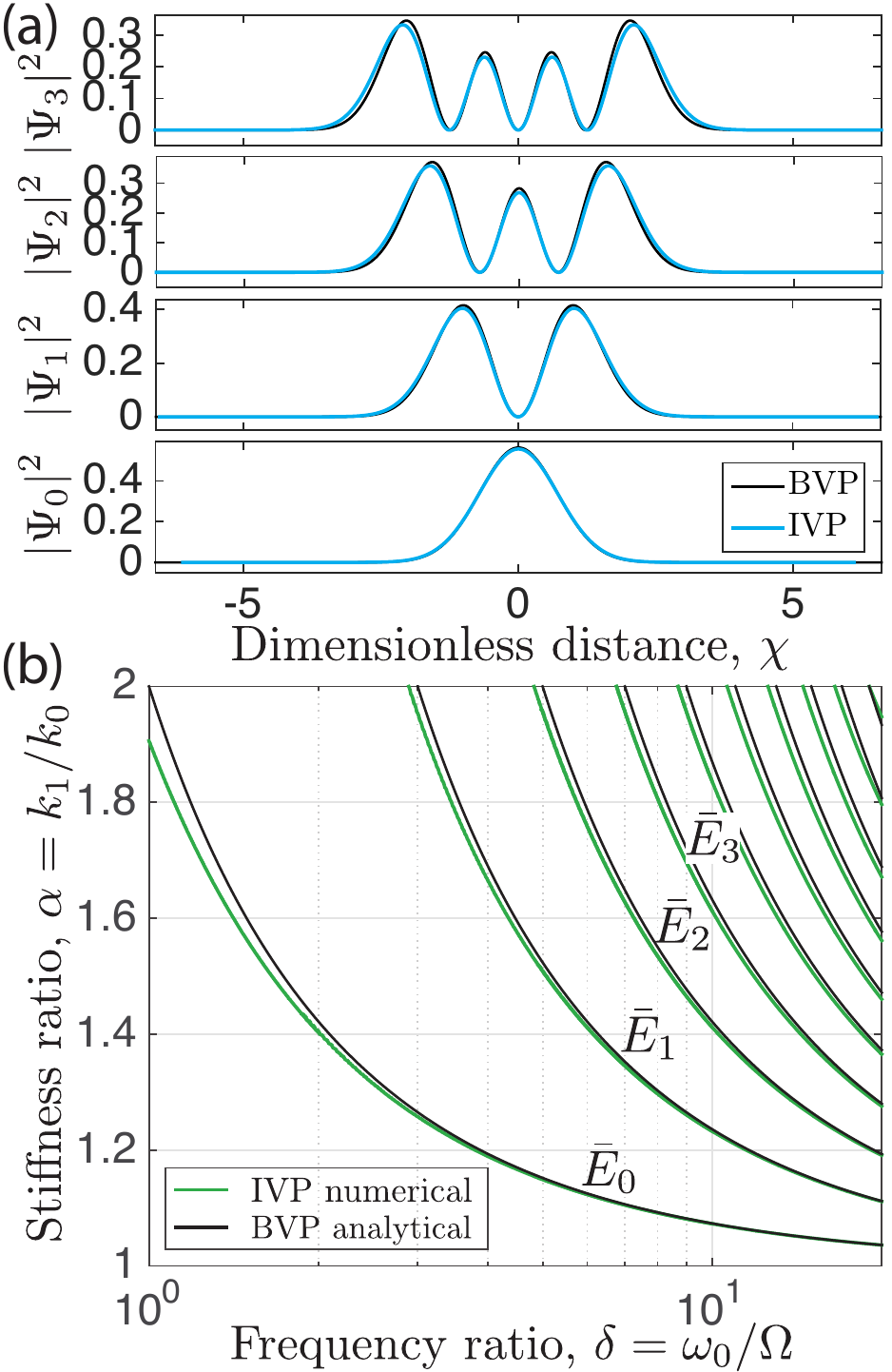}
\caption{Analytical predictions from a dimensionless Schr\"odinger equation with a harmonic potential. (a) Absolute square of the first four eigenfunctions $\Psi_l(\tau)$ on the dual primitive cell $-\hat{\chi}/2 \leq \chi < \hat{\chi}/2$ with $\chi = \bar{v}_{\theta}\tau$ and $\hat{\chi}=\bar{v}_{\theta}\bar{T}$. Black and blue lines are analytically and numerically obtained from Eq.(\ref{EqSchrod}) and Eq.(\ref{Eq1}) for $\delta=5$, respectively. The eigenfunctions are normalized so that 􏰃$N=\int_{\bar{T}}|\Psi(\tau)|^2d\tau =1$. (b) Limits of $\bar{T}$-periodic instability regions of Fig. \ref{Figure3}(a) in the $(\alpha,\delta)$ space for $1 \leq \alpha \leq 2$ and $1 \leq \delta \leq 20$. Numerical outcomes from the IVP Eq.(\ref{Eq1}) (green lines) are compared to the analytic results from the BVP Eq.(\ref{EqSchrod}) (black lines). Each solution $E_l=l+1/2$ of Eq.(\ref{EqEnergy}) corresponds to a branch of stability in the $(\alpha,\delta)$ space.}
\label{Figure7}
\end{center}
\end{figure}

\subsection{Reduction to a dimensionless Schr\"odinger stationary wave equation with a harmonic potential}

When reducing the regions of stable vibrational modes to the limit branches of periodic solutions $r_l(\tau) = \Psi_l(\tau)$ (that correspond to the stability frontiers $\sigma=0$ of the red regions in Fig. \ref{Figure3}(a)), $\Psi_l(\tau)$ is solution of the initial value problem (IVP), Eq.(\ref{Eq1}), so that we can write 
\begin{equation}
\frac{d^2\Psi(\tau)}{d \tau^2}-\Psi(\tau)+\alpha\cos(\tau /\delta)\Psi(\tau) = 0.
\label{Eqmotionpsi}
\end{equation}
\noindent
Unlike the general Eq.(\ref{Eq1}), Eq.(\ref{Eqmotionpsi}) is periodic and can therefore be reduced to a single representative period $\bar{T}$. The direct analytical analysis of Eq.(\ref{Eqmotionpsi}) is beyond the scope of this paper but one could simplify it thanks to numerical and physical observations. By construction and for sufficiently thin stability regions, the normalized periodic solutions $\Psi_l(\tau)$ can be decomposed in two exponential functions, mirrored with respect to a central vertical axis, connected by a function with $l+1$ extrema (see throughout Figs. \ref{Figure3}-\ref{Figure6}). As a consequence, the intensity of the normalized $\Psi_l(\tau)$ is localized in the center of the period and almost zero elsewhere (this is especially pronounced as $\delta$ increases and the width of the stability region $\epsilon$ decreases). So, multiplying $\Psi_l(\tau)$ by a $\bar{T}$-periodic function $F(\tau)$ would localized the intensity of $F(\tau)$ at the center of the period. Choosing the primitive periodic range $-\bar{T}/2 \leq \tau < \bar{T}/2$ as the representative cell for the whole solution $\Psi(\tau)$, $\Psi(\tau)\cos(\tau/\delta)$ would cancel the intensity of $\cos(\tau/\delta)$ away from the origin $\tau=0$ so that, assuming $\Psi(\tau)$ is localized enough, the Taylor series approximation 
\begin{equation}
\Psi(\tau)\cos(\tau/\delta) \approx \Psi(\tau)(1 - \tau^2/2\delta^2)
\label{Taylor}
\end{equation} 
\noindent
would be legitimate. Replacing Eq.(\ref{Taylor}) in Eq.(\ref{Eqmotionpsi}) and upon the change of variable $\chi=\bar{v}_{\theta}\tau$ with $\bar{v}_{\theta}=(\alpha/2\delta^2)^{1/4}$, Eq.(\ref{Eqmotionpsi}) governing $\Psi_l(\tau)$ can be rewritten in the form of a Boundary Value Problem (BVP) on the dual primitive cell $-\bar{v}_{\theta}\bar{T}/2 \leq \chi < \bar{v}_{\theta}\bar{T}/2 $
\begin{equation}
E\Psi(\chi)=\frac{1}{2} \chi^2\Psi(\chi) - \frac{1}{2}\frac{d^2\Psi(\chi)}{d \chi^2} 
\label{EqSchrod}
\end{equation}
where $E = \left(\left(\alpha-1\right)/\sqrt{2\alpha}\right)\times \delta$. The linear eigenvalue problem with variable coefficient in Eq.(\ref{EqSchrod}) is well-known as it is the dimensionless form of a stationary Schr\"odinger equation with a harmonic potential, predicting the total energy $E$ and wavefunction $\Psi(\chi)$ of a quantum harmonic oscillator \cite{Messiah1961}. For $\Psi(-\infty)=\Psi(+\infty)=0$, the discrete set of eigenvalues $E$ and eigenfunctions $\Psi(\chi)$ take the form 
\begin{equation}
E_l=\left(\left(\alpha-1\right)/\sqrt{2\alpha}\right)\times \delta=l+1/2
\label{EqEnergy}
\end{equation}
\noindent
and 
\begin{equation}
\Psi_l(\chi)=H_l(\chi)e^{(-\chi^2/2)}/(\pi^{1/4}\sqrt{2^ll!})
\label{EqWavefunctions}
\end{equation} 
\noindent
where $H_l(\chi)$ are Hermite polynomials and $l=0,1,2,\ldots$ The analytical results of Eqs.(\ref{EqEnergy})-(\ref{EqWavefunctions}) allow us to predict where will be the $l^{th}$ stability branch corresponding to a periodic vibrational mode with $\sigma=0$ in the modulation parameter space $(\alpha,\delta)$ as well as the shape of the periodic eigenfunctions $\Psi_l(\tau)$. Fig. \ref{Figure7}(a) illustrates, for $\delta = 5$, that the analytical eigenfunctions $\Psi_l(\chi)$ of Eq.(\ref{EqWavefunctions}) are in excellent agreement with the periodic eigenfunctions $\Psi_l(\tau)$ that were numerically computed from Eq.(\ref{Eq1}) and shown in Figs. \ref{Figure4}(a), (b) and (c) for a $\alpha$ in the $l=0$, $l=1$ and $l=2$ stability regions, respectively (note the results are shown as a function of $\chi$ but it could have been plotted as a function of $\tau$ as well). We have displayed the square of the absolute value of the eigenfunctions by analogy with the quantum harmonic oscillator. Fig. \ref{Figure7}(b) shows that each analytical $E_l$ of Eq.(\ref{EqEnergy}) accurately predict the $\bar{T}$-periodic limits of the $l^{th}$ numerical stability regions of Eq.(\ref{Eq1}) in the $(\alpha,\delta)$ space. As expected from numerical observations, since the width of the $l^{th}$ stability region and the errors in the approximation Eq.(\ref{Taylor}) decreases as $\delta$ increases, the analytical prediction Eq.(\ref{EqEnergy}) for each mode $l$ becomes better with $\delta$ as shown in Fig. \ref{Figure7}(b). 

Thus, in a certain asymptotic limit that will need to be rigorously defined in future work, the stationary wave equation Eq.(\ref{EqSchrod}) allows to compute, in the $(\alpha,\delta)$ modulation parameter space, all the eigenfunctions $\Psi_l(\tau)$ of the stable vibrational motion of the oscillator shown in Fig. \ref{Figure1}. As for the actual position of the mass of the oscillator, $r_l(\tau)$, it is not predicted in a classic deterministic way but it can be infer from this study that it will be a successive repetition of scaled periodic eigenfunctions. The successful so-called Copenhagen interpretation of quantum mechanics states that the quantity $|\Psi_l(\chi)|^2$ of the $l^{th}$ mode of the quantum oscillator represents the probability to find the particle at a position $\chi$ \cite{Messiah1961}. If to eventually pursue the analogy with the presented mechanical oscillator, one would have no choice but to introduce two supplementary concepts: (i) a measurement concept already discussed in the previous section and extended in Appendix A that could justify a probabilistic interpretation, (ii) a relativity and periodicity concept, maybe of the same nature that the original work of De Broglie \cite{DeBroglie1924}, that would fix the oscillator of Fig. \ref{Figure1}(b) on a periodically moving Galilean frame $\mathcal{R}'$ with constant velocity $v_{\theta}$ in order to relate time $\tau$ to a finite space $-\bar{v}_{\theta}\bar{T}/2 \leq \chi < \bar{v}_{\theta}\bar{T}/2$.

\section{Conclusions and discussions}
In summary, we presented a theoretical study of an overlooked fundamental mechanism: a harmonically modulated $1D$ linear oscillator whose mass is naturally collapsing. We have shown that the physics of the stable vibrational modes of this system is analogous in time to the $1D$ quantum harmonic oscillator (QHO). In an asymptotic limit in the modulation parameter space, the presented system behave like an effective classic harmonic oscillator, a situation similar to the so-called correspondence principle \cite{Messiah1961}. In the opposite limit, the stable modes of vibration are quantized in thin stability regions in the modulation space. Those neutrally stable modes can be quantitatively described by the time-energy uncertainty relation of the QHO when using an analogy with the statistical interpretation of quantum mechanics.  Finally, in the limit where the stability regions are thin, we observed a behavior reminiscent of a wave-particle duality: the original initial value problem (IVP) governing the motion of the particle can be reduced to a boundary value problem (BVP) in a primitive periodic cell that is a dimensionless Schr\"odinger stationary wave equation with a harmonic potential. The solutions of Schr\"odinger's equation represent the motion of the particle albeit with an ignorance of the causality of this motion which could be reconstruct by statistical means.

This paper is a first attempt, mainly numerical, to study an overlooked fundamental mechanism analogous to the QHO. In future work, a more rigorous mathematical approach would be needed notably to characterize the asymptotic limit where the original equation of motion can be approximated by a dimensionless Schr\"odinger's equation. Meanwhile, this theoretical work raises an important question: could this modulated oscillator be a reality? An answer would be to set up an experimental system whose linearized equation of motion is in the form of Eq.(\ref{Eq1}). The simplest realization could be an inverted pendulum whose pivot point is vertically vibrated which is the archetypal example of a mass that naturally collapses under periodically modulated gravity. The major problem with such a basic experiment would lie in the theoretical width of the stability regions that we uncovered. The latter are intrinsically so small that any additional experimental noise would probably made the pendulum unstable. A possibility could be to add some damping to the oscillator that would decrease the growth rate $\Re(\sigma)$ and widen the instability regions for any given modulation parameters. The drawback of his approach is that the eigenfunctions $\Psi(\tau)$ as well as the actual position $r(\tau)$ would loose their symmetry in the primitive periodic cell. Another solution would be to start the experiment from modulation parameters for which the stability regions are reasonably large and strive to progressively explore the regions as it shrink. In any case, a study of the dependence of an additional noise on the stability of the motion in a nonlinear framework would be needed to get some physical insights in the transitions between stable regions. 

Despite the aforementioned practical difficulties, the presented dynamical system already refines the classical picture of linear oscillators and sheds new light on periodically time-varying systems governed by Floquet theory that are at the heart of many problems in physics. Its appealing mechanical analogy with a fundamental quantum system as well as its relative simplicity call for further theoretical and experimental explorations for an impact that would be twofold. It could be seen as a pedagogical dynamical system that can help explaining the complexity of the concepts of quantum physics. But for the most audacious, it could be regarded as a very first brick to eventually pave the way to a new realistic interpretation of quantum mechanics. The current scientific content is by any means too restricted to foresee any future implications in that sense but considering our mechanical analog is based on particular representative vibrational modes that are non-local nor causal entities, at least one can hope since it may not fault Bell's famous no-go theorem that forbid any hidden variables theory to ever explain quantum phenomena.

\begin{figure}[!t]
\begin{center}
\includegraphics[width=0.82\columnwidth]{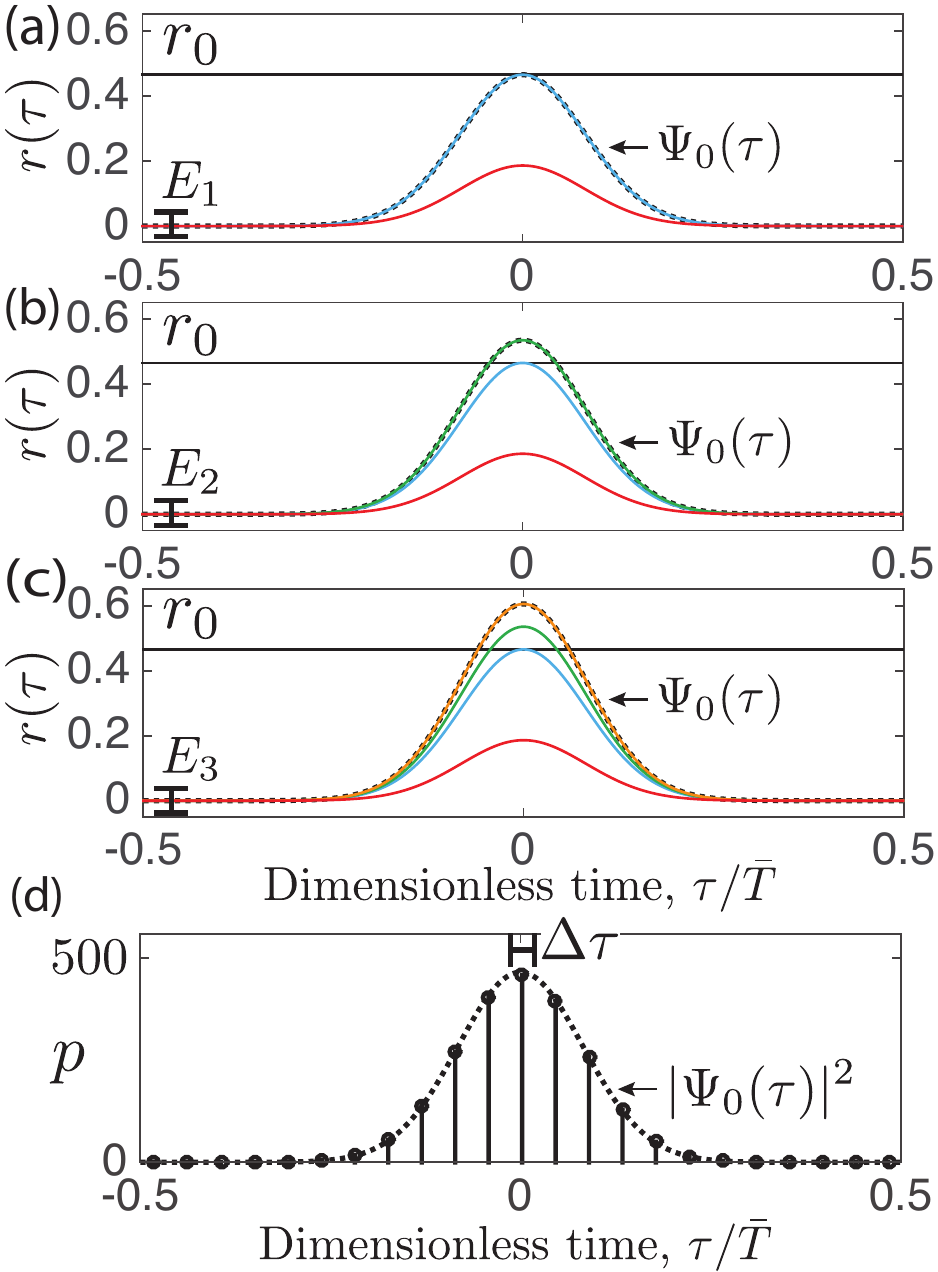}
\caption{Schematic illustration of the time-energy uncertainty relation found in Eq.(\ref{Equncertainty}). (a) Eigenfunction $\Psi_0(\tau)$ in dotted lines and two random positions $r(n\bar{T}-\bar{T},n\bar{T})$ in full lines for $E(r(0))=1/2\dot{r}(0)^2+1/2(-1+\alpha)r(0)^2=E_1$. (b) Same plot as (a) but for an initial energy $E_2=E_1+\epsilon$. (c) Same plot as (b) but for an initial energy $E_3=E_2+\epsilon$. (d) Statistical vision showing the number of times $p$ the particle has been detected at time $\tau$ over a number $P$ of independent measurements. The dotted line shows the square of the modulus of the periodic eigenfunction, $|\Psi_0(\tau)|^2$.}
\label{Figure8}
\end{center}
\end{figure}

\section*{ACNOWLEDGEMENTS}

The author is grateful to O. Devauchelle, F. James, C. Josserand, R. Lagrange, P.Y. Lagr\'ee, E. Lajeunesse, Y. Privat, P. M. Reis, M. Rossi and D. Terwagne for fruitful discussions.

\section*{APPENDIX}

\subsection*{Statistical interpretation and measurement problem}

\vspace{0cm}
Assume a modulated linear oscillator of Fig. \ref{Figure1}(b) for $\delta >> 1$ so that no direct numerical simulations or measurement of the motion of the particle $r(\tau)$ is possible because the width of the stability regions is so small. Eventually, the only way to get informations on that system is from the neutrally stable eigenfunctions $\Psi_l(\tau)$ predicted by the Schr\"odinger equation Eq.(\ref{EqSchrod}). Inspired by the presented study, we could model our ignorance on the actual position $r(\tau)$ by assuming the latter is a succession of randomly scaled eigenfunctions $\Psi_l(\tau)$, between $-1$ and $1$. Assume we have a measuring device that, because of its precision limitations and sensitivity, could detect the presence of the particle but only above a triggering amplitude of motion, $r(\tau)=r_0$. One could possibly reconstruct the $\Psi_l(\tau)$ from indirect measurements and statistical techniques.

Figs. \ref{Figure8}(a)-(c) show the evolution of the eigenfunction $\Psi_0(\tau)$, normalized to $N=\int_{\bar{T}}|\Psi(\tau)|^2d\tau =1$, in the representative periodic cell $-\bar{T}/2 \leq \tau < \bar{T}/2$, for $\delta = 5$. If we assume we have been able to make a measurement, Fig. \ref{Figure8}(a) represents the worst case scenario where the initial energy $E(r(0))=1/2\dot{r}(0)^2+1/2(-1+\alpha)r(0)^2=E_1$ is such that the maximum of the eigenfunction $\Psi_0(\tau)$ is at $r_0$. For this given initial energy $E_1$, the only possibility is to measure that the particle will be at the center $\tau=0$ of the representative period, i.e. the center of any possible period (this situation corresponds to the full blue line). Any other possible situation, illustrated for example by the red line, could not be measured. For some very small relative fluctuations of the initial conditions, i.e. of the initial energies $E_2$ and $E_3$ (where $E_3 > E_2$), the amplitude $r(\tau)$ can be greatly affected in the center of the period because of the divergent nature of $\Psi_l(\tau)$ as shown in Figs. \ref{Figure8}(b) and (c). In those cases, for one independent measurement, the device could detect a particle before $\tau = 0$ which is the case for $r(n\bar{T}-\bar{T},n\bar{T})$ in green or orange lines. But for another measurement, the $r(n\bar{T}-\bar{T},n\bar{T})$ represented by a blue line could arrive first in which case the particle will be detected again in $\tau=0$. 

In the end, if we perform for a fluctuation of initial energies, a large set of $P$ independent measurements to obtain the number of time $p$ the particle has been detected in $\tau$, we could obtain the probability plot in a representative period $-\bar{T}/2 \leq \tau < \bar{T}/2$ given in Fig. \ref{Figure8}(d). The probability to find the particle at $\tau$ should eventually take the same shape than the square of the eigenfunction modulus $|\Psi_0(\tau)|^2$ because the more intense $\Psi_0(\tau)$, the higher the chance will be that $r(\tau)$ is large. Furthermore, the uncertainty $\Delta_{\hat{E}}$ on the initial energy $E(r(0))$ would be related to the uncertainty $\Delta_{\tau}$ on $\tau$. If we knew precisely the initial energy, we would still have a finite uncertainty on $\tau$ because of the random scaling process and vice-versa, it is not because the location in time $\tau$ is detected with certainty that the initial energy is unique. Note that the same reasoning could be made for modes in the $l>0$ stability branches with the same expected result that the density of probability to detect the particle in $\tau$ could follow $P(\tau)=\Psi^{\ast}(\tau)\Psi(\tau)=\left|\Psi(\tau)\right|^2$. It would be interesting in future work to perform an actual statistical study of the problem described in Fig. \ref{Figure8}.

\subsection*{Frequency spectrum and sensitivity}

\begin{figure}[!h]
\begin{center}
\includegraphics[width=0.82\columnwidth]{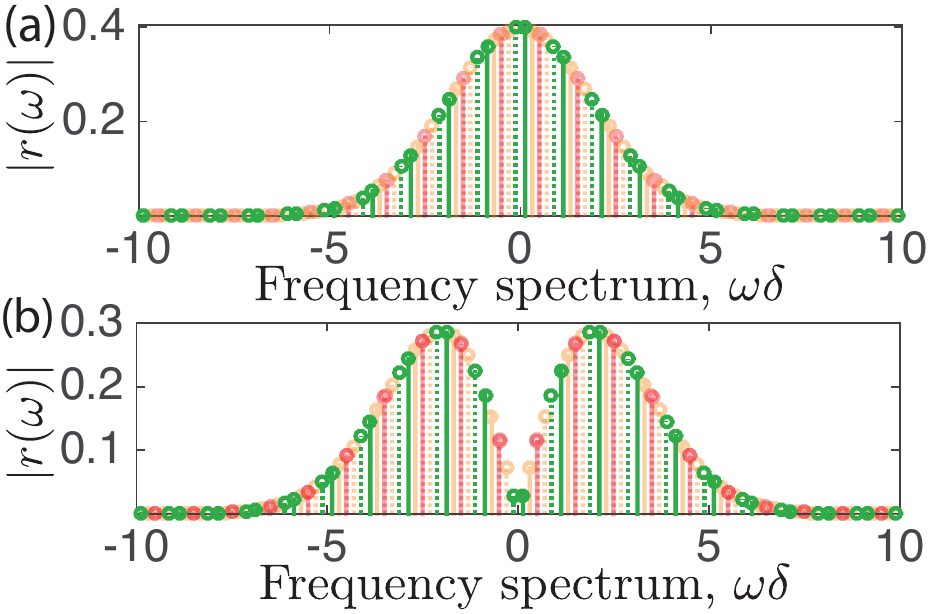}
\caption{Discrete Fourier transforms of the stable vibrational modes of Figs. \ref{Figure5}(a) and (b) for $\delta = 5$. (a) Mode in the stability region $l=0$ with $\alpha = 1.14905152323442$. (b) Mode in the stability region $l=1$ with $\alpha = 1.9549540063305$. Dotted and solid lines represent the Floquet form $\Psi^{\ast}(\tau)e^{-i\Im(\sigma)\tau}$ and $\Psi(\tau)e^{i\Im(\sigma)\tau}$, respectively. Various colors correspond to a perturbation of the parameter $\delta$ with an $\epsilon$ of the order of $10^{-13}$ so that $\Im(\sigma) \approx 0.25/2\delta$, $0.62/2\delta$ or $0.98/2\delta$.}
\label{Figure9}
\end{center}
\end{figure}
\vspace{0cm}

Another sensitivity property of the stable vibrational modes with respect to $(\alpha,\delta)$, of the same nature of the one shown in Section \ref{reduc1} in the time domain, can be retrieved in the frequency domain. Figs. \ref{Figure9}(a) and (b) show the frequency spectrum of the computed solutions $r(\tau)$ displayed in Figs. \ref{Figure5}(a) and (b), for $\delta=5$ and a $\alpha$ in the stable region $l=0$ and $l=1$, respectively. Various colors correspond to a slight perturbation $\epsilon$ of $\delta$ so that the fundamental frequency $\Im(\sigma)$ reads, $\Im(\sigma) \approx 0.25/2\delta$, $\Im(\sigma) \approx 0.62/2\delta$ and $\Im(\sigma) \approx 0.98/2\delta$. For each color, full and dotted lines represent a Floquet form $\Psi^{\ast}(\tau)e^{-i\Im(\sigma)\tau}$ and $\Psi(\tau)e^{i\Im(\sigma)\tau}$, respectively. Whatever the fundamental frequency $\Im(\sigma)$, a stable vibrational mode in the $l^{th}$ stability region is characterized by a poly-harmonic symmetric spectrum $r(\omega)$ with $(l+1)$ local maxima. The spectrum is sensitive to $\Im(\sigma)$ and therefore to the modulation parameters in the sense that it can be shifted inside the cell $0 \leq  \Im(\sigma) \leq 1/2\delta$ by slightly varying $(\alpha,\delta)$ as shown by the various colors in Fig. \ref{Figure5}. However, the global shape and notably the energy spectral density of the mode, $\bar{E} = \frac{1}{2\pi}\sum|r(\omega)|^2$, is insensitive to $(\alpha,\delta)$ since it varies from one stability region to another, very much alike the eigenfunction $\Psi_l(\tau)$ or the limits of the spectral cell  $0 \leq  \Im(\sigma) \leq 1/2\delta$. Finally, it is interesting to note that, upon scaling in the $x$ and $y$-axis, the square of the modulus of the eigenfunctions $|\Psi_0(\tau)|^2$ and $|\Psi_1(\tau)|^2$ would perfectly envelope the discrete Fourier spectrum shown in Figs. \ref{Figure9}(a) and (b), respectively.

\providecommand{\noopsort}[1]{}\providecommand{\singleletter}[1]{#1}%


\begin{thebibliography}{27}
\expandafter\ifx\csname natexlab\endcsname\relax\def\natexlab#1{#1}\fi
\expandafter\ifx\csname bibnamefont\endcsname\relax
  \def\bibnamefont#1{#1}\fi
\expandafter\ifx\csname bibfnamefont\endcsname\relax
  \def\bibfnamefont#1{#1}\fi
\expandafter\ifx\csname citenamefont\endcsname\relax
  \def\citenamefont#1{#1}\fi
\expandafter\ifx\csname url\endcsname\relax
  \def\url#1{\texttt{#1}}\fi
\expandafter\ifx\csname urlprefix\endcsname\relax\def\urlprefix{URL }\fi
\providecommand{\bibinfo}[2]{#2}
\providecommand{\eprint}[2][]{\url{#2}}


\bibitem{Guckenheimer1983}
J. Guckenheimer and P. Holmes, \textit{Nonlinear oscillations, dynamical systems, and bifurcations of vector fields} (NewYork Springer Verlag 1983).
\bibitem{Strogatz2001}
S.H. Strogatz, \textit{Nonlinear dynamics and chaos: with applications to physics, biology and chemistry} (Perseus publishing 2001).
\bibitem{Nakamoto1986}
K. Nakamoto, \textit{Infrared and Raman spectra of inorganic and coordination compounds Part A} (John Wiley \& Sons 2009).
\bibitem{Nayfeh2008}
A. H. Nayfeh and P.F. Pai, \textit{Linear and nonlinear structural mechanics} (John Wiley \& Sons 2008).
\bibitem{berge1984}
P. Berg\'e, Y. Pomeau, and C. Vidal, \textit{Order within chaos} (John Wiley \& Sons 1984).
\bibitem{bolotin1964}
V.V. Bolotin, \textit{The dynamic stability of elastic systems} (Holden- Day, lnc 1964). 
\bibitem{Turner1998}
K.L. Turner, S.A. Miller, P.G. Hartwell, N.C. MacDonald, S.H. Strogatz, and S.G. Adams, Nature \textbf{396}, 149 (1998).
\bibitem{Amin2012}
M. A. Amin, R. Easther, H. Finkel, R. Flauger, and M.P. Hertzberg, Phys. Rev. Lett. \textbf{108}, 241302 (2012).
\bibitem{Kumar1994}
K. Kumar and L.S. Tuckerman, J. Fluid Mech. \textbf{279}, 49 (1994).
\bibitem{Melo1994}
F. Melo, P. Umbanhowar, and H.L. Swinney, Phys. Rev. Lett. \textbf{72}, 172 (1993).
\bibitem{Engels2007}
P. Engels, C. Atherton, and M.A. Hoefer, Phys. Rev. Lett. \textbf{98}, 095301 (2007).
\bibitem{Hill1886}
G. W. Hill, Acta mathematica, \textbf{8}, 3-36 (1886).
\bibitem{Poincare1886}
H. Poincar\'e, Bulletin de la soci\'et\'e math\'ematique de France, \textbf{14}, 77-90 (1886).
\bibitem{Lazarus2010}
A. Lazarus, B. Prabel, and D. Combescure, J. Sound Vib. \textbf{329}, 3780 (2010).
\bibitem{Bastidas2012}
V.M. Bastidas, C. Emary, B. Regler, and T. Brandes, Phys. Rev. Lett. \textbf{108}, 043003 (2012).
\bibitem{Kapitsa1951}
P.L. Kapitsa, Uspekhi Fiz. Nauk \textbf{44}, 7 (1951).  
\bibitem{Grescho1970}
P.M. Grescho and R.L. Sani, J. Fluid Mech. \textbf{40}, 783 (1970).
\bibitem{Stephenson1908}
A. Stephenson, Philos. Mag., \textbf{15}, 233 (1908).
\bibitem{Acheson1993}
D. Acheson and T. Mullin, Nature, \textbf{366}, 215 (1993).
\bibitem{Messiah1961} 
A. Messiah, \textit{Quantum Mechanics, Volume 1} (North-Holland 1961).
\bibitem{Merali2015}
Z. Merali, Nature \textbf{521}, 278 (2015).
\bibitem{DeBroglie1924}
 L. De Broglie, Philos. Mag., \textbf{47}, 446-458 (1924).
\bibitem{Bohm1952}
D. Bohm, Phys. Rev., \textbf{85}, 166 (1952).  
\bibitem{Protiere2006}
S. Proti\`ere, A. Boudaoud, and Y. Couder, J. Fluid Mech. \textbf{554}, 85 (2006).
\bibitem{Couder2006}
Y. Couder and E. Fort, Phys. Rev. Lett., \textbf{97}, 154101 (2006).
\bibitem{Bush2015}
J. W. M. Bush, Phys. Today \textbf{68}, 47  (2015).
\bibitem{Bush2015b}
J.W.M. Bush, Annu. Rev. Fluid Mech. \textbf{47} (2015).
\bibitem{Eddi2009}
A. Eddi, E. Fort, F. Moisy, and Y. Couder, Phys. Rev. Lett. \textbf{102}, 240401 (2009).
\bibitem{Fort2010}
E. Fort, A. Eddi, A. Boudaoud, J. Mukhtar, and Y. Couder, Proc. Natl. Acad. Sci. USA, \textbf{107}, 17515 (2010).
\bibitem{Perrard2014}
S. Perrard, M. Labousse, M. Miskin, E. Fort, and Y. Couder, Nat. Commun. \textbf{5}, 17515 (2014).
\bibitem{Wilczek2012}
F. Wilczek, Phys. Rev. Lett., \textbf{109}, 160401 (2012).
\bibitem{Zhang2017}
J.P. Zhang et al., Nature, \textbf{543}, 217-220 (2017).
\bibitem{Choi2017}
S. Choi et al., Nature, \textbf{543}, 221-225 (2017).
\bibitem{Sacha2017}
K. Sacha and J. Zakrzewski, arXiv preprint arXiv:1704.03735 (2017).
\bibitem{Whittaker1996} 
E. T. Whittaker and G. N. Watson, \textit{A course of modern analysis} (Cambridge university press 1996).
\bibitem{Magnus2013}
W. Magnus and S. Winkler, \textit{Hill's equation} (Courier corporation 2013).
\bibitem{Floquet1879}
G. Floquet, Annales scientifiques de l'\'Ecole Normale Sup\'erieure \textbf{8}, 3 (1879).
\bibitem{Moore2005}
G. Moore, SIAM J. Numer. Anal. \textbf{42}, 2522 (2005).
\bibitem{Lazarus2010b}
A. Lazarus and O. Thomas, Comp. Rend. Acad. Sci.: Mecanique \textbf{338}, 510 (2010).
\bibitem{Lazarus2017}
B. Bentvelsen and A. Lazarus, under review (2017).
\bibitem{Movies}
See Supplemental Material at [URL will be inserted by publisher] for movies showing the simulations of time evolutions of a naturally unstable particle on a harmonically modulated potential energy hill for various modulation parameters $(\alpha,\delta)$. The first movie shows $\alpha=1.8$ and various $\delta > 1$ when the second movie shows $\delta = 3.78861850488$ and $\alpha>1$. 
\bibitem{DeBroglie1924}
L. De Broglie, L., Philos. Mag. \textbf{47}, 446-458 (1924).


\end{thebibliography}
\end{document}